# EFFECTS OF NEUTRON RADIATION ON THE THERMAL CONDUCTIVITY OF HIGHLY ORIENTED PYROLYTIC GRAPHITE

M.A. Guazzelli[1], L.H. Avanzi[1], V.A.P. Aguiar[2], A.C. Vilas-Bôas[1], S.G. Alberton[2], S.H. Masunaga[1], E.F. Chinaglia[1], K. Araki[3], M. Nakamura[3], M.M. Toyama[3], F.F. Ferreira[4], M. T. Escote[4], R.B.B. Santos[1], N.H. Medina[2], J.R.B. Oliveira[2], F. Cappuzzello[5,6], M. Cavallaro[6] for the NUMEN collaboration

[1] Centro Universitário da FEI, São Bernardo do Campo, São Paulo, Brazil
[2] Instituto de Física, Universidade de São Paulo, São Paulo, Brazil
[3] Instituto de Química da Universidade de São Paulo, São Paulo, SP, Brazil
[4] Universidade Federal do ABC, Santo André, São Paulo, Brazil
[5] Dipartimento di Fisica e Astronomia "Ettore Majorana", Università di Catania, Catania, Italy
[6] Istituto Nazionale di Fisica Nucleare–Laboratori Nazionali del Sud, Catania, Italy

**Abstract**. Highly Ordered Pyrolytic Graphite (HOPG) has been extensively researched due to its chemical and physical properties that make it suitable for applications in several technologies. Its high thermal conductivity makes HOPG an excellent heat sink, a crucial characteristic for manufacturing targets used in nuclear reactions, such as those proposed by the NUMEN project. However, when subjected to different radiation sources, this material undergoes changes in its crystalline structure, which alters its intended functionality. This study examined HOPG sheets before and after exposure to a 14 MeV neutron beam. Morphological and crystallographic analyses reveal that even minor disruptions in the high atomic ordering result in modifications to its thermal properties. The results of this study are essential to establish the survival time of the HOPG used as thermal interface material to improve heat dissipation of a nuclear target to be bombarded by an intense high-energy heavy-ion beam.

## 1. Introduction

This research focuses on crucial factors such as defining the suitable materials to be used as substrates for thin film targets, their specifications and properties, and the effective dissipation of the substantial heat generated during irradiation with high-energy ions. Once the backing (substrate) for the nuclear reaction target is determined, another essential aspect is determining the duration for which the composite target (reaction element + heat dissipation backing) can be continuously irradiated. A significant challenge arises from the high heat production during nuclear reactions, which can lead to the melting of target materials, particularly those with low melting points, rendering the study of reaction products impossible. Various proposals suggest the use of highly oriented pyrolytic graphite (HOPG) as a backing for depositing thin film targets, offering the desired functionality for the composite target.

HOPG, composed of stacked graphene sheets, is known for its exceptional in-plane thermal properties, contributing to efficient heat dissipation. It is anticipated that these properties will play a crucial role in facilitating effective heat removal during experiments, thus helping to prevent target melting. However, the outstanding thermal conductivity of HOPG heavily relies on the perfection of its crystalline structure, which becomes compromised due to the numerous defects induced by irradiation [1, 2, 3]. As the crystalline structure of HOPG deteriorates, its thermal conductivity decreases and may eventually approach that of amorphous carbon. Before reaching that stage, it becomes necessary to replace the target material [4,5].

### 1.1 The NUMEN Project

These composite targets will be used in the NUMEN (NUclear Matrix Elements for Neutrinoless double beta decay) project. The primary objective of NUMEN is to develop an innovative experimental approach for determining the nuclear matrix elements essential in calculating the half-life of neutrinoless double beta decay ($0\nu\beta\beta$) across a wide range of systems. The project focuses on investigating double charge exchange (DCE) reactions induced by heavy ions, which exhibit intriguing similarities to $0\nu\beta\beta$ due to the involvement of the same initial and final nuclear states. By studying DCE reactions, the project aims to extract crucial information about the nuclear matrix elements and their influence on the $0\nu\beta\beta$ decay process [5-10].

Furthermore, the NUMEN project aims to explore the competition between the direct DCE mechanism and multinucleon transfer processes by simultaneously measuring other relevant reaction channels. This comprehensive investigation will enable a deeper understanding of the underlying physics and provide insights into the interplay between different reaction mechanisms. By advancing our knowledge of the nuclear matrix elements and their impact on neutrinoless double beta decay, the NUMEN project may contribute to the broader nuclear and particle physics field. The findings from this project will enhance our understanding of fundamental processes in the universe and contribute to ongoing efforts in neutrino research [5,7].

### 1.2 The HOPG

This research revolves around the crucial aspects of defining suitable target materials, their specifications, properties, substrate deposition, and effective heat dissipation for the significant amounts of heat generated in nuclear reactions. Since highly oriented pyrolytic graphite (HOPG) is established as the support material for the nuclear reaction target, determining the uninterrupted duration of the composite target's irradiation becomes essential [4, 11-14].

Highly oriented pyrolytic graphite is composed exclusively of carbon atoms arranged in a hexagonal shape, making it a lamellar material. Its crystalline structure consists of parallel and stacked layers, exhibiting a remarkable degree of three-dimensional ordering. It shares similarities with graphene, which is organized in a two-dimensional form. Essentially, HOPG can be described as a sequence of stacked graphene layers [14-16].

The substantial heat produced during nuclear reaction irradiation can cause the target material to melt, making the study of the reaction products impossible. Due to its excellent in-plane thermal properties, HOPG finds applications in many areas, such as electronics, aerospace, and energy storage. The combination of high thermal conductivity and mechanical flexibility makes HOPG a promising material for thermal management in flexible devices. Therefore, it is also used as a thermal interface material to improve heat dissipation in devices, as in the case of a target substrate for nuclear reactions, allowing better performance and reliability. The excellent thermal conductivity of HOPG relies heavily on the perfection of its crystalline structure, but an extensive number of defects (such as point vacancies, bivacancies, and Stone-Wales) induced by irradiation can compromise it. As the crystalline structure of

HOPG deteriorates, its thermal conductivity decreases, and it may eventually approach that of amorphous carbon. Before reaching this limit, it becomes necessary to replace the target material [17,18]. Simulations using SRIM/TRIM (Transport of ions in matter [19]) allowed us to estimate the effect of the density of defects in the HOPG crystal lattice on its heat dissipation capacity [4,19].

### 1.2.1 – HOPG Thermal Conductivity

Despite its highly ordered and layered structure, HOPG exhibits anisotropic thermal conductivity, meaning it conducts heat differently along different crystallographic directions [13,14]. The thermal conductivity of HOPG can reach values as high as 2000-3000 W/m.K in the plane of the carbon atoms, making it one of the most efficient conductors of heat among known materials. This remarkable thermal conductivity arises from the strong covalent bonding within each graphene layer [2].

In accordance with Reference [13], significant variations in the thermal conductivity of carbon manifest across its diverse allotropic configurations. Amorphous carbon exhibits a thermal conductivity range of 0.01 to 2.0 W/m.K within the temperature range of 0 to 500 K. In contrast, the in-plane pyrolytic graphite demonstrates thermal conductivity values ranging from 900 to 1000 W/m.K at the same temperature interval, peaking at approximately 5000 W/m.K around 100 K. At room temperature (300 K), amorphous carbon registers a thermal conductivity of approximately 1.5 W/m.K, while cross-plane pyrolytic graphite and polycrystalline graphite exhibit values of 5.0 W/m.K and 200 W/m.K, respectively. Notably, HOPG showcases an on-plane thermal conductivity of 1700 W/m.K at 300 K, underscoring the susceptibility of its thermal conductivity to subtle alterations in crystallinity.

Gaining a comprehensive understanding of the properties of graphene, and consequently HOPG, in comparison to different materials, is crucial for defining their technological applications. One of the techniques employed to achieve this understanding is the utilization of classical molecular dynamics simulations. These simulations show that even a mere 0.25% of defects in graphene lead to a significant 50% reduction in its thermal conductivity [14]. Therefore, we predict that HOPG may have similar behavior.

This study aimed to establish a characterization methodology to define a correlation between irradiation-induced defects and their impact on the physical and chemical properties of HOPG. In particular, commercial thin HOPG sheets (1.0 x 1.0 cm$^2$) with a 2.0 µm thickness were irradiated with a monoenergetic neutron generator of 14 MeV in three different exposure times. The samples were characterized before and after irradiation by using Atomic Force Microscopy (AFM), Raman spectroscopy, X-ray diffraction, Scanning Electron Microscopy (SEM), Electrical Resistivity, Magnetoresistance, and Thermal Conductivity measurements [3]. As demonstrated below, structural defects lead to a loss in HOPG's ability to conduct heat, which is crucial for applications such as target support for nuclear reactions proposed in the NUMEN project.

## 2.- Materials and Methods

### 2.1 - Neutron Irradiation procedure

In order to make an experimental exploration of the effects of fast neutron irradiation on commercial HOPG foils (grade ZYA, from Optigraph GmbH), the Deuteron-Tritium (D-T) neutron generator (Thermo Scientific™ MP 320) at *Instituto de Estudos Avançados (*IEAv) in São José dos Campos, Brazil was used. This generator produces a monoenergetic neutron beam of ~14 MeV. The typical neutron yield provided by this generator is approximately $10^8$ n/s. The energy of the neutron beam was accurately measured using a 100 µm-thick fully depleted Si surface barrier (SSB) charged particle detector, which detected the $^{28}$Si(n, α)$^{25}$Mg, and $^{28}$Si(n, p)$^{28}$Al nuclear reactions.

The analysis considered that, to determine the effects of neutron irradiation, most of the impact would result from the capture of neutrons by the $^{12}$C nuclei and the subsequent nuclear processes involving the compound nucleus. Cross sections of the fusion and evaporation of different particles were calculated using the PACE4 (Projection Angular momentum Coupled Evaporation) program. PACE4 is a statistical model evaporation code that employs Monte Carlo simulation to de-excite the compound nucleus [20]. The compound nucleus, $^{13}$C, can de-excite by emitting gamma-rays or evaporating particles. In the latter case, it can evaporate one neutron leaving a recoiling $^{12}$C nucleus, or evaporate an alpha particle (4He nucleus), leaving a recoiling $^9$Be nucleus. Although every condition can lead to lattice disorders, the cross-section for the $^{13}$C recoil is about two orders of magnitude below the other processes and will not be considered in the following analysis. Table I shows the production cross-section of the by-products of this nuclear reaction, which are α particles and $^9$Be isotopes and excited $^{12,13}$C nuclei, are shown. All the by-products of this nuclear reaction are accountable for inducing damage to the material's crystalline structure. These nuclear reactions with 14 MeV neutrons can be observed in reference [21].

**Table I:** Yelds of residual nuclei from the n + $^{12}$C reaction at 14 MeV.

| Isotope | Cross section (mb) | Angular range of emission (degree) | Energy range (MeV) |
|---|---|---|---|
| $^{13}$C[a] | 2.71 | | |
| $^{12}$C | 250 | 0 – 36 | 0 – 3 |
| $^9$Be | 162 | 0 – 36 | 2 – 4 |
| $^4$He (with $^9$Be) | 162 | 0 – 180 | 2 – 8 |

[a] The cross-section for $^{13}$C is negligible.

### 2.2 - TRIM Simulations

TRIM software was employed to estimate the quantities of vacancies generated in HOPG due to the recoil of $^{12}$C and $^9$Be residual nuclei and the emitted alpha particle in the latter [19]. TRIM is a Monte Carlo simulation program embedded in the SRIM package, which calculates ion energy loss, scattering, range, target sputtering yield, vacancy formation, and implanted ion distribution through binary collision calculations, considering the universal ion-atom potential. This code makes it possible to assess the concentration of vacancies created by the charged particle impact on a target [19]. To obtain a detailed description of the number and distribution of vacancies induced by reaction products, an input file of $10^5$ particles with the same proportions and angle/energy distributions generated by the PACE4 code, as shown in Table I, was used. As the probability of interaction of the neutrons along the target does not change, the position of the recoiling nuclei and alpha particles inside HOPG was evenly distributed [1].

We did not consider the potential damage to the crystalline structure of HOPG caused by $^{13}$C isotopes since its production cross-section is significantly smaller than other residual nuclei. Figure 1(a) shows the vacancies distribution per primary reaction products and recoiling lattice atoms capable of creating damage cascades as a function of depth within the HOPG due to 14 MeV neutrons. Figure 1 also shows energy transfer to the lattice, 1(b), and energy transfer to ionization of target atoms, 1(c), as a function of depth within the HOPG due to evaporation products and lattice recoiling atoms [1, 2, 3]. It is important to note that, as mentioned in the literature, these three mechanisms can induce modification of carbon bonding, though the processes involved may not be the same [22-24].

It is important to note, however, that the calculations using TRIM are not a complete picture of the complex phenomena involved, as TRIM does not take into account local heating (as in thermal spike model – ref. [25]) through electron-phonon coupling; only consider phonon generation by ions without energy enough to produce new displacements; and, in the case of HOPG, it also does not consider that ionization can also break sp2 bonds, thus altering Raman spectra. Atom dislocations caused by ions are known to promote graphene-layer mixing, as presented in reference [26], who used molecular dynamics (MD) simulations to calculate these effects. However, MD was not used in this work due to the computational effort to simulate such a complex ensemble of high-energy particles in a material as thick as 2 micrometers.

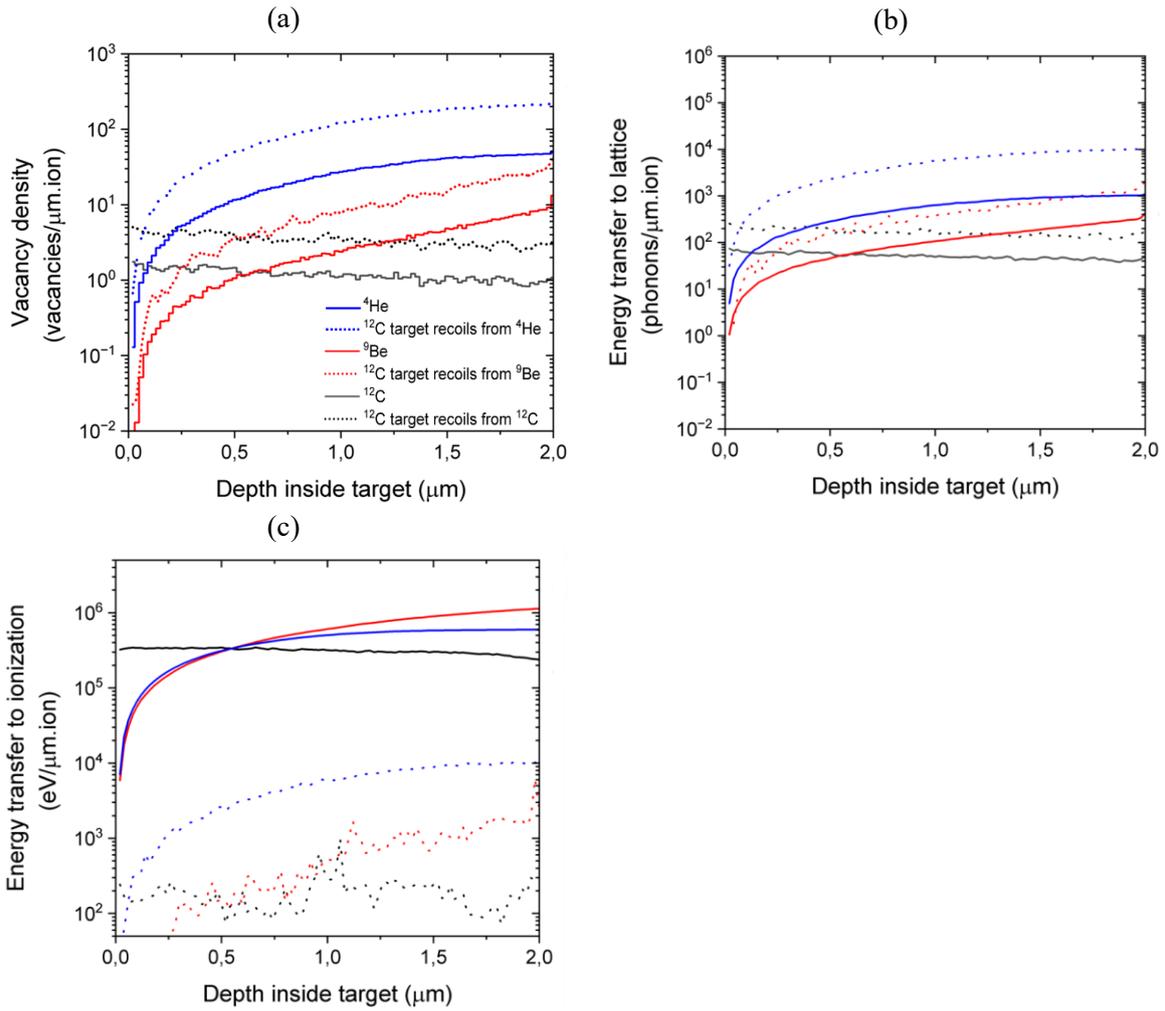

*Figure 1: Distribution of vacancies per ion (a), energy transfer to lattice (b), and energy transfer to ionization of target atoms (c) as a function of depth within the HOPG due to evaporation products and lattice recoiling atoms following reaction 14 MeV neutrons with $^{12}C$.*

The experimental exploration in this work focuses on three 1.0 cm$^2$ HOPG foils, positioned at about $(11.30 \pm 0.25)$ cm far from a D-T neutron generator, which were exposed to different fluences of fast neutron bombardment. Table II presents the irradiation time and neutron fluence incident on each HOPG foil. The irradiated samples will be compared with a virgin HOPG foil (pristine).

Table II - Irradiation time and neutron fluence incident on each HOPG foil.

| SAMPLE | Irradiation Time (hours) | Fluence (n/cm$^2$) |
|---|---|---|
| N2 | 10.5 | 5.18(13)×10$^9$ |
| N3 | 24.4 | 1.27(2)×10$^{10}$ |
| N4 | 30.5 | 1.7(3)×10$^{10}$ |

It is essential to point out that during the NUMEN experiments, there will be a production of neutrons estimated at 10$^5$ neutrons/cm$^2$/s [27], with the total spectrum consisting of thermal and epithermal neutrons, ranging from 10$^{-3}$ eV to 10$^8$ eV. This study used monochromatic neutrons of $1.4 \times 10^7$ eV, which may represent a portion of the neutrons produced in the reaction.

The orientation and disorder of the HOPG crystal lattice of these three foils and a pristine foil were characterized using Raman spectroscopy and X-ray Diffraction (XRD). The microstructural characterizations were done with scanning electron microscopy (SEM) and atomic force microscopy (AFM). These characterizations are presented in the following sections.

**2.3 – Characterization of orientation and disorder in the crystal lattice - XRD and Raman characterization**

High-resolution X-ray diffraction data were collected on a STADI-P diffractometer (Stoe®, Darmstadt, Germany), operating in transmission geometry at 40 kV and 40 mA, using Cu$K$a$_1$ ($\lambda$ = 1.54056 Å) radiation filtered by a primary beam monochromator [Ge (111) curved crystal], equipped with a 0.5 mm divergence slit and a 3 mm circular scattering slit and a silicon strip detector (Mythen 1 K, from Dectris®, Baden, Switzerland). The measurements were performed in the angular range from 25.000º to 93.250º (2θ) with steps of 0.015º and integration time of 180 s at each 1.05º. With this setup, changes in full width in the half maximum (FWHM) of the diffraction peaks can be ascribed only to the effects of irradiation on the HOPG crystal lattice. In this configuration (transmission geometry), and due to the HOPG´s highly oriented crystal structure, the only reflection found in the measurements for samples before and after irradiation were the ones shown in Figure 2. It is worth mentioning that conventional measurements, in reflection geometry, would reveal the basal planes parallel to the surface, i.e., the (00$l$) planes. On the other hand, in transmission geometry, we observe peaks perpendicular to the sample plane ($h$00), an effect previously observed by other authors [28].

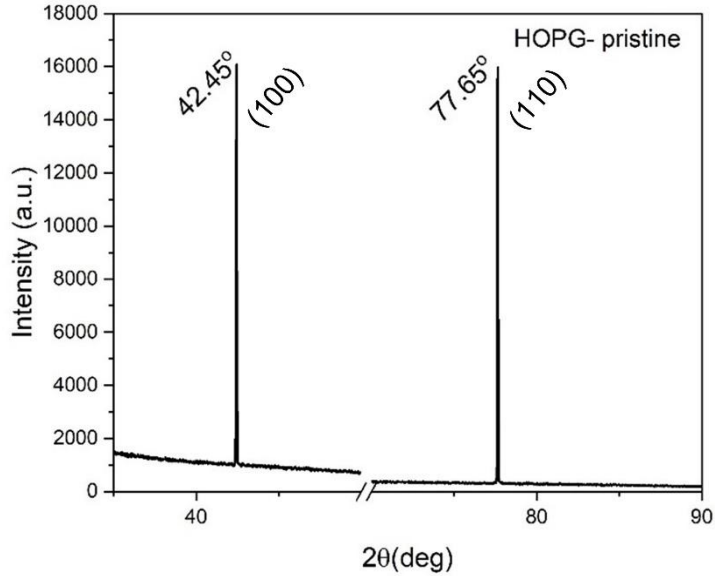

***Figure 2:*** *X-ray diffractogram (θ-2θ) showing only reflections of planes perpendicular to the sample surface. This result indicates a high ordering of atoms in the HOPG crystal lattice.*

In materials science and nanotechnology, micro-Raman analysis has proven to be an invaluable tool for characterizing the structural and chemical properties of various materials. This non-destructive spectroscopic method enables researchers to gain insight into the atomic and molecular composition of materials, offering valuable information on their crystallinity and structural changes.

Confocal Raman microscopy measurements were performed with a WITEC alpha 300R equipment using a Nd:YAG laser ($\lambda$ = 532 nm) and a beam diameter of 800 nm. The penetration depth of the green laser in HOPG is approximately 50 nm, ensuring that the Raman spectra predominantly capture information from a thin, near-surface layer of the sample. Spectral images were obtained by scanning the sample in the x,y direction with a piezo-driven xyz feed-back controlled scan stage and collecting a spectrum at every pixel. All experiments were conducted with a Nikon objective (100 × NA = 0.8), applying a laser power of 30 mW/cm$^2$ on the sample stage. Raman images were generated from the band height intensities measured at the maximum of the peaks. In this study, micro-Raman mapping analysis was employed to investigate the effects of neutron beam exposure on a set of samples. About 15 different areas of each sample were carefully examined, allowing a comprehensive evaluation of their properties. Each mapping was performed with a 10x10 μm$^2$ area at a resolution of 35x35 pixels (286 nm pixel size) with an integration time of 1 second on each pixel. One of the key parameters studied was the D/G ratio, a fundamental metric in Raman spectroscopy that provides critical information about the degree of disorder and crystallinity in a material.

### 2.4 – Electrical and Thermal transport

One can assess the crystallinity in highly crystallized graphite materials by examining two key parameters: the residual resistivity ratio (RRR) and the maximum transverse magnetoresistance $(\Delta\rho/\rho)_{max}$ [29, 30]. RRR is the ratio of the in-plane electrical resistivity at room temperature ($\rho_{RT}$) to that at low temperatures, such as 10 K ($\rho_{10K}$). The transverse magnetoresistance $(\Delta\rho/\rho)_{max}$ can be determined by

$$\left(\frac{\Delta\rho}{\rho}\right)_{max} = \frac{\rho(H) - \rho(0)}{\rho(0)} \quad (1)$$

where $\rho(H)$ is the electrical resistivity obtained under applied magnetic field $H$, and $\rho(0)$ denotes the resistivity measured in the absence of $H$.

Electrical resistance ($R$) as a function of temperature ($T$) and applied magnetic field was conducted utilizing a standard four-probe method on a 9-T Physical Properties Measurement System (PPMS) manufactured by Quantum Design. The orientation of the current and applied magnetic field was aligned parallel and perpendicular, respectively, to the HOPG basal plane.

The steady-state thermal conductivity κ(T) measurements were taken using a homemade apparatus based on the parallel heat-flow technique [31,32]. Rectangular HOPG samples (typically 0.2 μm x 4 mm x 6 mm) were glued to this apparatus with silver epoxy, where the upper and low edges of the sample were fixed to the hot and cold terminals of the sample holder. All measurements were performed at room temperature (300 K) inside a close-cycle cryostat (ARS -system) at high vacuum (~$10^{-5}$ mbar).

### 2.5 - Microstructural characterization - SEM and AFM

Atomic Force Microscopy (AFM) and Scanning Electron Microscopy (SEM) analyses allow the correlation of changes induced in the physical properties of HOPG due to the damage produced in its crystal structure by irradiation [15,17,33].

Based on the well-organized structure of carbon atoms constituting the graphene layers within HOPG, we anticipate that the presence of vacancies and atomic displacements resulting from the irradiation process can be identified through changes in the roughness and topography profiles of the examined samples, along with a reduction in stacking orientation [14].

The HOPG surface morphology was analyzed before and after each neutron beam exposure time with atomic force microscopy using a Shimadzu SPM9700. Images were acquired in the air in dynamic mode using commercial Si tips. The samples were analyzed in 4 different scan sizes (0.5 μm to 6.0 μm) with an average of 14 images per sample at aleatory places. Since the RMS (root mean square) surface roughness depends on the observation scale, it was obtained from the average of at least four images acquired in different areas of the HOPG with scan areas of (1.0 x 1.0) μm$^2$ and (2.0 x 2.0) μm$^2$. The interlayer step distance analysis was obtained using (2.0 x 2.0) μm$^2$ images.

Scanning electron microscopy (SEM) was conducted across various regions of each of the four highly oriented pyrolytic graphite (HOPG) foils to comprehensively observe potential alterations in their microstructure resulting from defects induced by exposure to a fast neutron beam. The microscope used for the study is the Mira 3 XMU model from Tescan, Field-Emission Gun (FEG), with a resolution of 1.0 nm at 30 kV. This instrument allows us to obtain detailed, high-precision images, enabling comprehensive exploration of nanoscale features and structures.

## 3 – Results and discussions

### 3.1 - XRD Analysis

Figure 3 shows the X-ray diffractogram in the transmission geometry for all samples. From the plots, it was possible to obtain the FWHM and the lattice parameters variation for all HOPG samples, and the results are in Table III. To accurately model peak shapes, we required at least five data points spanning from the left to the right side of each peak, taking the full width at half maximum (FWHM) as a reference. Given that our equipment operates with a fixed step size of 0.015º, capturing a peak with sufficient data points necessitated measuring each diffraction pattern three times. For each measurement, the pattern was shifted by 0.005º initially and subsequently merged to ensure comprehensive data coverage across the peak profile. The lattice parameter c for sample N4 showed no variation. The peak intensities depend on instrument parameters, preferred orientation, and thermal displacement of the atoms in the structure. This is further affected by the presence of defects, such as those introduced during neutron irradiation [30]. As the integrated intensity of a diffraction peak depends on structural, specimen, and instrumental factors, the former depends on the atomic crystal structure, related to the relative positions of atoms within the unit cell [34]. As HOPG crystallizes in a hexagonal crystal system with space group $P6_3mc$ (nr. 186 in the International Tables for Crystallography vol. A), the carbon atoms occupy Wyckoff sites 2a (0, 0, $z$; $z = 0$) and 2b (1/3, 2/3, $z$; $z = 0.005$). On the other hand, we could not refine their fractional coordinates (as indicated above, only the "z" value is allowed to vary since the "x" and "y" values of the atomic positions cannot change without breaking the symmetry of the specified unit cell) as we observed only two independent reflections in the measured diffraction pattern. The intensity loss with increased irradiation is attributed to defect formation that is represented by an increase in the atomic displacement parameter in the refinements.

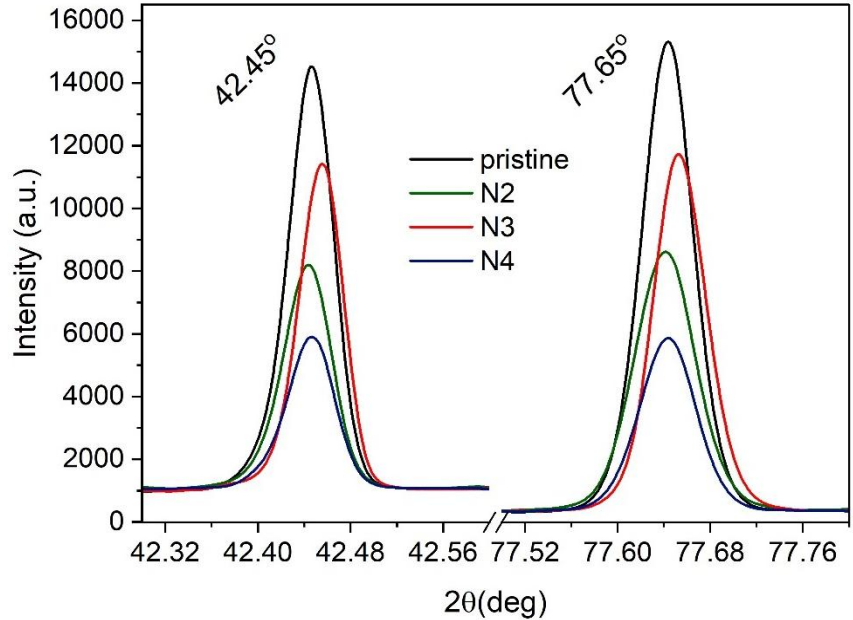

*Figure 3:* X-ray diffractograms for HOPG pristine and irradiated N2, N3, and N4 samples.

Table III: FWHM and lattice parameters variation for all HOPG samples

| SAMPLE | FWHM (deg) | $\Delta a/a$ | $\Delta c/c$ |
|---|---|---|---|
| Pristine | 0.0348 (4) | ($a_o$=2.4582 Å) | ($c_o$=6.681 Å) |
| N2 | 0.0376 (8) = 8.0% | 4.07 (65) × $10^{-5}$ | 7.48 (49) × $10^{-4}$ |
| N3 | 0.0335 (7) = - 3.7 % | 12.20 (12) × $10^{-5}$ | 7.48 (27) × $10^{-4}$ |
| N4 | 0.0368 (9) = 5.7% | 8.14 (18) × $10^{-5}$ | 0 |

From Figure 3, it is possible to observe a decreasing tendency in the peak intensities for irradiated samples, and, from the data analysis, slight variations were also determined in lattice parameters *a* and

*c* and in FWHM. However, these variations are within the estimated standard deviations inferred for the unit cell parameters. The light variations are justified by the low fluence of neutrons. However, even a small variation in the crystalline perfection of HOPG can lead to a significant variation in its thermal conductivity, as can be seen in reference [14].

### 3.2 - Raman Spectroscopy

The Raman spectrum is renowned for its high sensitivity to structural changes, making it a powerful tool for material characterization, particularly for those with a high degree of crystallinity, enabling the detection of subtle alterations in the crystal lattice. The typical spectrum of an HOPG sample exhibits a well-defined G-band at approximately 1580 cm$^{-1}$. The presence of a D-band is associated with the degree of disorder within the structure [35,36]. Given that the relative intensity between the two peaks varies with the degree of disorder in HOPG, the ratio between IG (intensity of the G-band) and ID (intensity of D-band) enables the estimation of the disorder evolution as a result of the damage caused by irradiation effects. ID and IG are the intensity of the D and G peaks in the normalized spectrum, respectively [36].

Figure 4 presents the average Raman spectra from the regions within the Raman maps where the D band was detected for each of the four analyzed samples. It is noticeable that the D band intensities, associated with the presence of defects, increase with the exposure time of the HOPG foil to the neutron beam. This suggests a degradation in the thermal properties of the material.

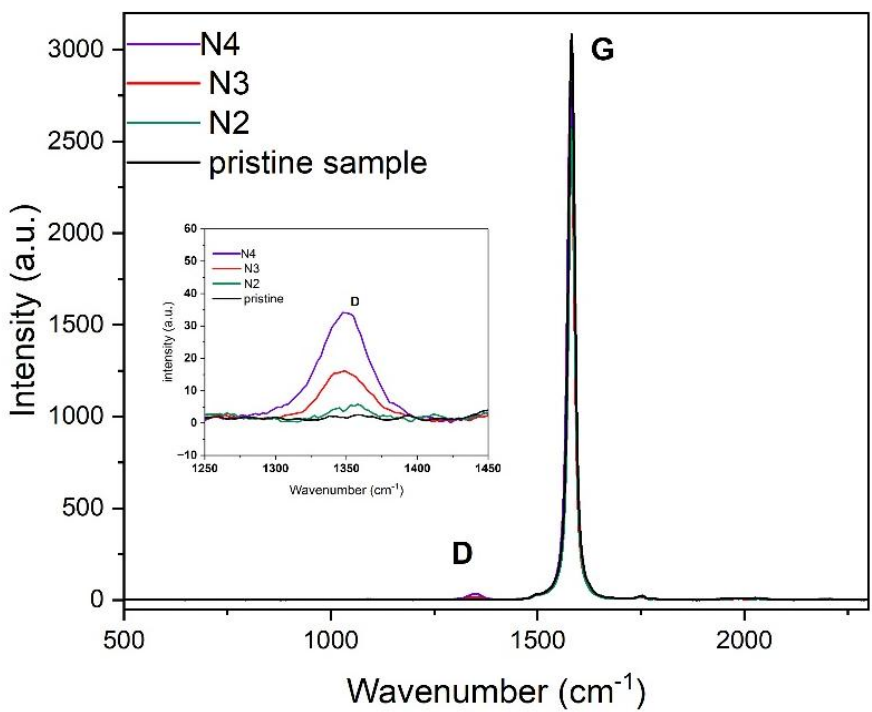

***Figure 4:*** *The micro-Raman spectra for HOPG pristine and irradiated: N2, N3 and N4 samples. The spectra represent the average Raman spectra of the regions in the Raman maps where the D band was observed.*

This work's statistical analysis considered the intensity of the first two peaks referring to the D and G bands. The images presented in Figure 5 serve as illustrative representations of one of the 10x10 μm² regions that underwent analysis within each HOPG specimen following exposure to fast neutron irradiation. The histograms represent the distributions of the intensity ratios of the D/G bands in the area analyzed by Raman mapping. It is important to note that the histograms for the D/G band ratio are derived solely from mapping points where a D-band signal was detected. This selective approach ensures that the analysis remains focused on capturing relevant structural variations.

Despite the sample before being exposed to the neutron beam already presenting a small number of defects, represented by the D/G = 0.0035(8) ratio, the results of this analysis reinforce that, as the exposure time to the neutron beam increases, the D/G ratio also increases. The samples exhibit a subtle but noticeable loss of crystallinity [37]. This gradual increase, totaling 163% in the D/G ratio, indicates that the structural order of the materials was undergoing a transformation. This phenomenon can be attributed to the interaction of neutrons with the atomic structure of the samples, leading to structural modifications and a decrease in the degree of crystallinity.

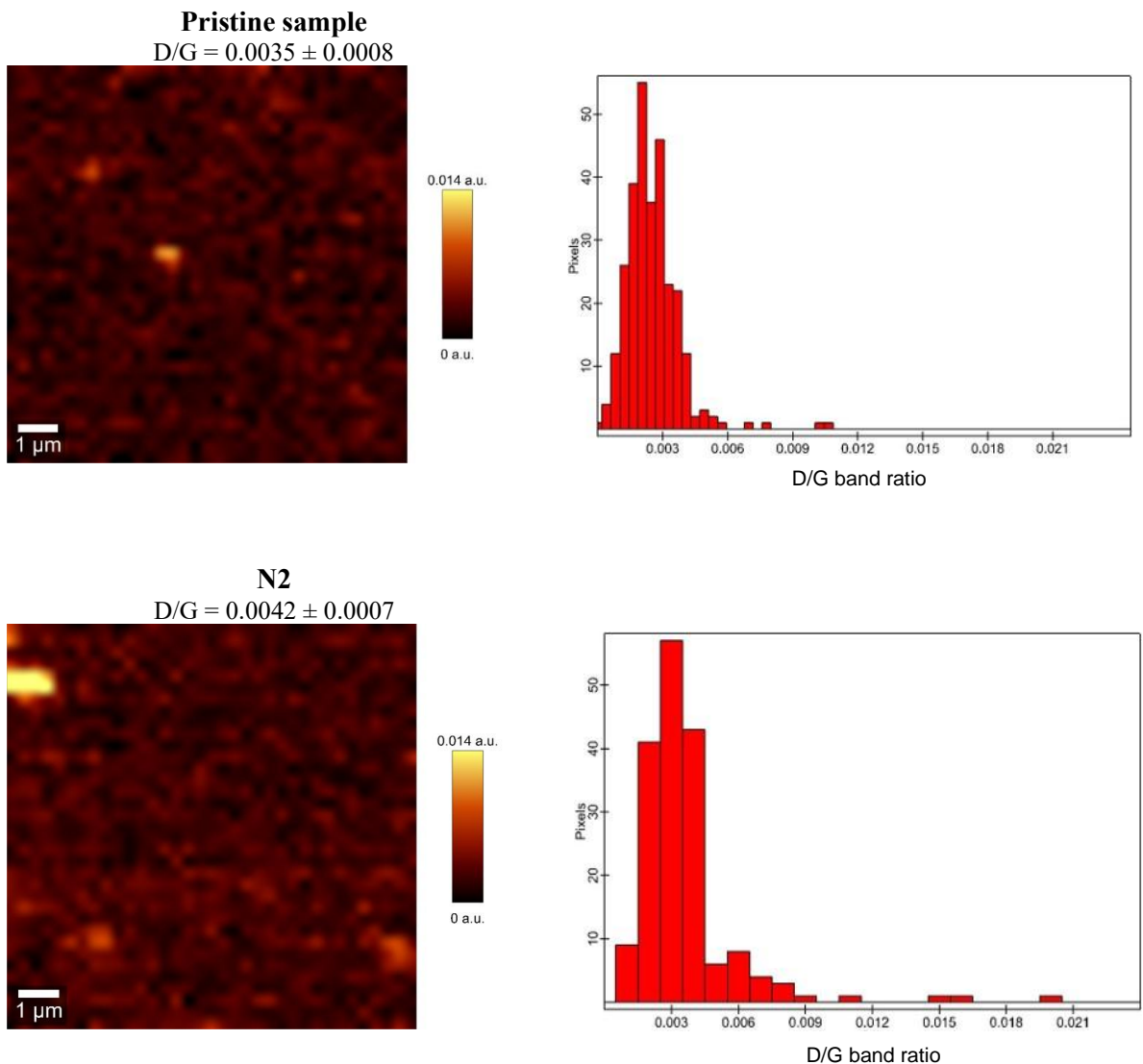

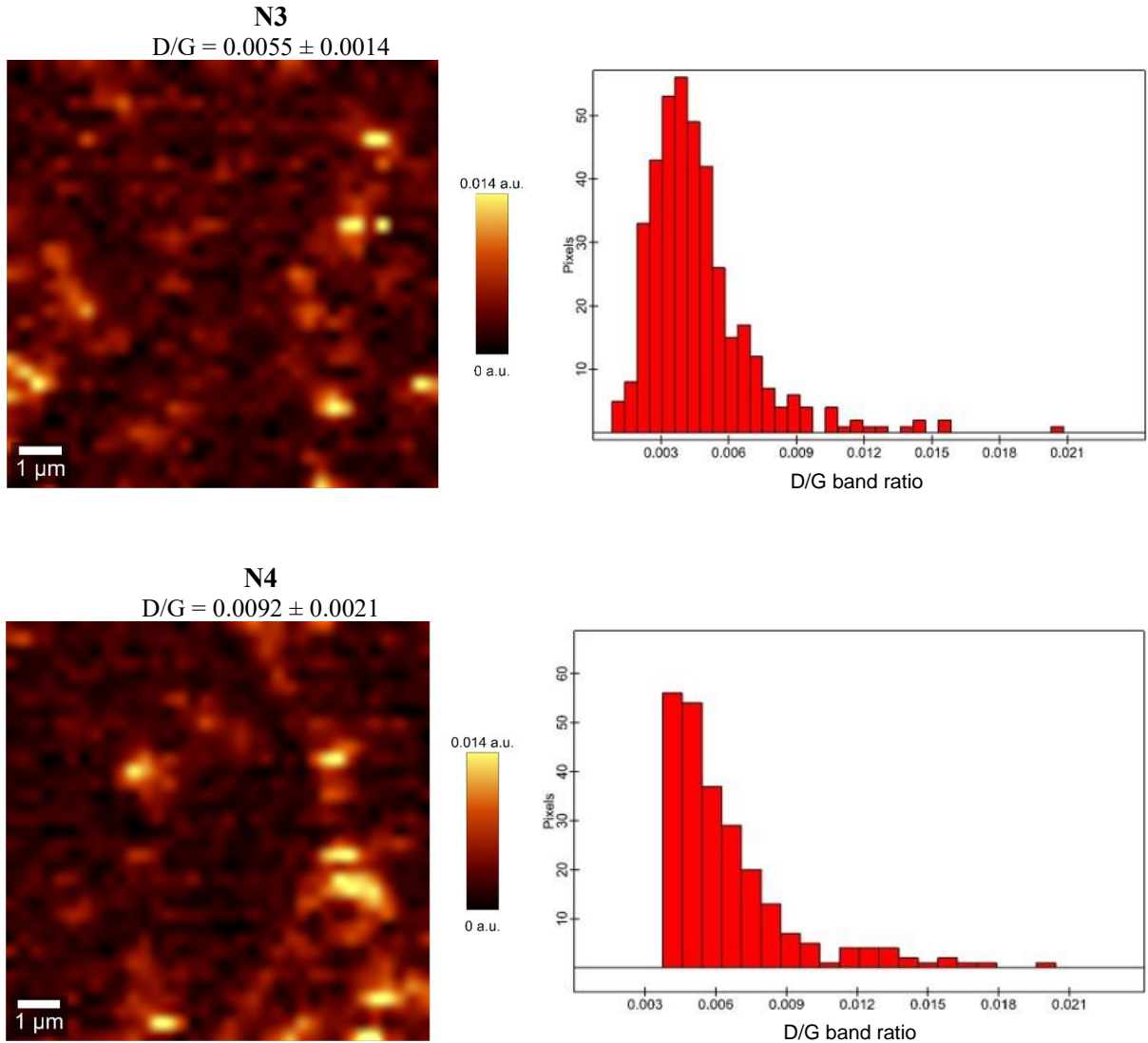

*Figure 5:* *Representative micro-Raman images of the 10x10 μm² regions for pristine, N2, N3, and N4 with their respective D/G peaks intensity ratio. The histograms represent the distributions of the D/G band intensity ratios in the area analyzed.*

### 3.3 Electrical Resistivity and Magnetoresistance

Figure 6(a) shows the in-plane electrical conductivity ($\sigma$) of pristine HOPG sample as a function of temperature. The $\sigma$ value at room temperature (RT) aligns with the manufacturer's provided value of $2.1 \cdot 10^6$ $(\Omega \cdot m)^{-1}$, as indicated by the red dashed line on the graph. The inset of Figure 6(a) displays the residual resistivity ratio, which serves as an indicator of the crystalline quality of the materials under investigation [30]. The RRR values decrease from 8.3 for the pristine sample to 5.8, 5.7, and 3.4 for irradiated samples N2, N3, and N4, respectively. This decline suggests a reduction in the degree of crystallinity for samples N2, N3, and N4 exposed to the neutron beam, corroborating the findings of Raman spectroscopy analysis.

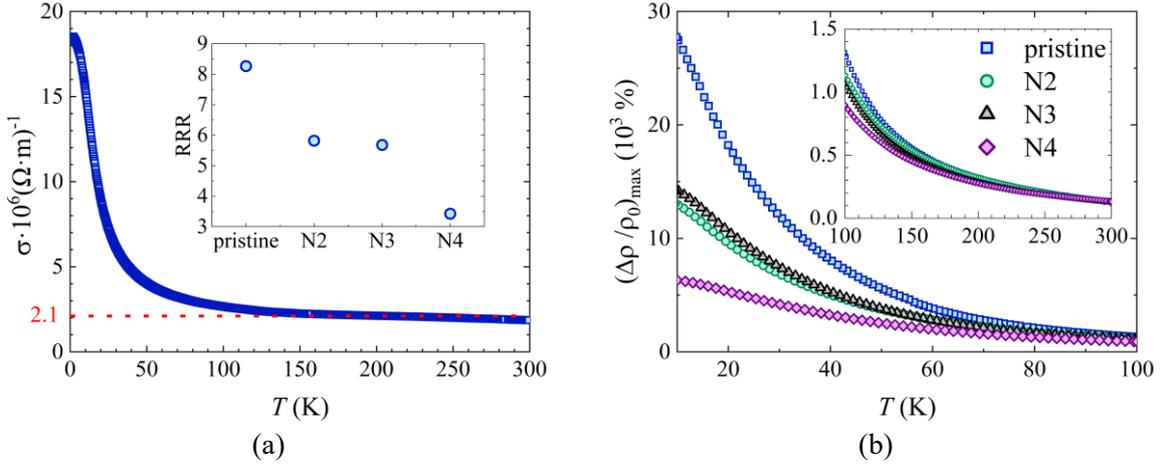

*Figure 6: (a) Temperature dependence of electrical conductivity for the pristine HOPG foil. The dashed red line indicates the manufacturer-provided value of σ at RT. The inset presents the residual resistivity ratio for the pristine HOPG foil alongside irradiated foils N2, N3 and N4.
(b) Magnetoresistance was observed under an applied magnetic field of 1 T for the pristine HOPG sample, and samples N2, N3, and N4.*

The temperature-dependent maximum magnetoresistance, $(\Delta\rho/\rho)_{max}$, measured under an applied magnetic field of 1 T, is shown in Figure 6 (b). At low temperatures, defect scattering has a significant influence on electrical resistivity, thus resulting in notable variations in the $(\Delta\rho/\rho)_{max}$ relative to crystal perfection. For instance, at 10 K, $(\Delta\rho/\rho)_{max}$ values for samples N2 and N3 diminish to 45% and 49%, respectively, of the maximum magnetoresistance observed for the pristine sample. In contrast, the $(\Delta\rho/\rho)_{max}$ curves for the pristine HOPG, as well as samples N2, N3, and N4, demonstrate convergence near RT, as illustrated in the inset of Figure 6 (b).

Table IV shows the estimated in-plane thermal conductivity κ(300K) values for pristine HOPG, N2, N3, and N4 samples at room temperature. The pristine sample exhibits κ(300 K) ≈ 1249(62) W/m.K, close to the 1700 W/m.K reported for this commercial HOPG foil. The irradiated samples also display a remarkable decrease in the thermal conductivity values to 511(31), 602(16), and 568(11) W/m.K compared to the pristine sample. Such a κ(T) reduction of more than 50% is related to the increased defect observed in these samples; as described by Raman results, these samples show modification and loss of crystallinity due to neutron irradiation. The thermal conductivity of graphite is predominantly governed by phonon transport, and due to the large interlayer spacing and weak interlayer bonding forces, the phonon dispersion in graphite exhibits a quasi-two-dimensional nature [38]. This leads to significantly higher thermal conductivity within the basal plane. However, this conductivity is highly sensitive to grain size and defects within the crystal lattice [39]. Lower values of κ indicate structurally compromised material, where κ is constrained by phonon scattering at grain boundaries, point defects, dislocations, and other structural imperfections [13]. In addition, our findings are consistent with those reported in reference [38], which showed that the fragmentation of basal planes—resulting from prolonged neutron and electron irradiation—also contributes to the observed reduction in thermal conductivity.

**Table IV**: In-plane thermal conductivity κ(T) measured at 300 K for all samples.

| SAMPLE | κ(T) (W/m.K) | κ(T) decrease (%) |
|---|---|---|
| Pristine | 1249 (62) | - |
| N2 | 511 (31) | 59.1 |
| N3 | 602 (16) | 51.8 |
| N4 | 568 (11) | 54.2 |

Figure 7 shows behaviors of in-plane thermal conductivity (300 K) and transverse magnetoresistance (10 K) as a function of neutron fluences in which, as expected, a significant reduction in the values of this parameter is observed, a behavior explained by the increasing number of defects produced in the HOPG crystal lattice, as indicated in reference [13]. This behavior suggests a correlation between electron and phonon transport properties with the increase of the amount of the defect in these samples. Although both quantities present this similar reduction tendency, they are measured in different temperatures, one at 300 K and the other at 10 K, and it is challenging to relate any connection between the spins and lattice phonons in pristine and irradiated HOPG samples. However, this point should be intensively exploited in future works with complementary thermal conductivity measurements as a function of temperature.

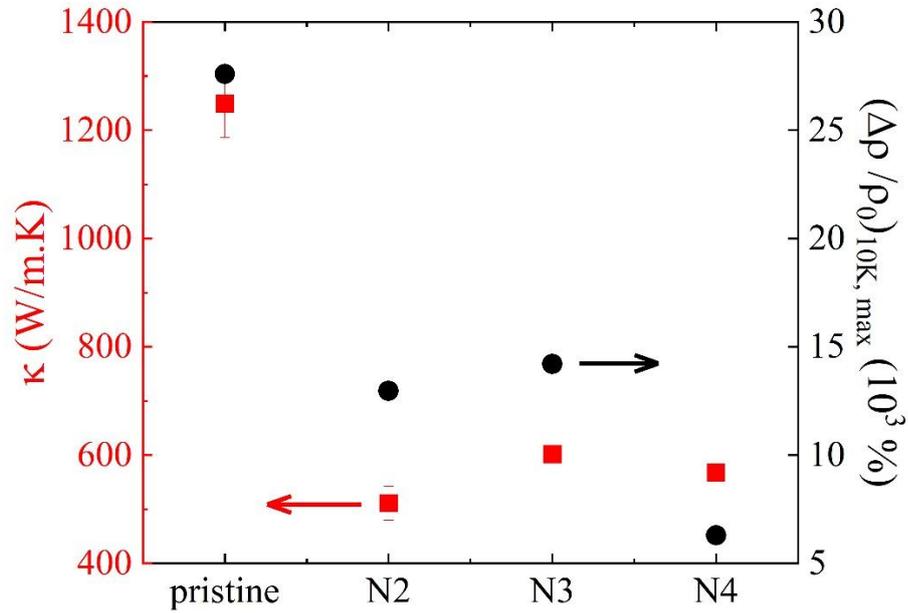

*Figure 7:* *Thermal conductivity and transverse magnetoresistance as a function of neutron fluences.*

### 3.4 - Scanning Electron Microscopy and Atomic Force Microscopy

The scanning electron microscopy (SEM) and atomic force microscopy (AFM) images representing the typical behavior of each of the four HOPG foils are shown in Figure 8. In these images, it is possible to clearly see that after the neutron irradiation process, the HOPG layers are loosely bound in the cross-plane, indicating the bond is weaker and the layers tend to be peeled off. These findings support the quantitative results obtained through XRD and Micro Raman techniques.

In the comparison between the pristine sample and the one subjected to prolonged neutron irradiation, N4, more pronounced evidence of degradation in the stacking of graphene foils comprising the HOPG, becomes apparent. Regions of loosely arranged layers overlapping the sample surface were observed, contrasting with the previously homogeneous layer structure. The illustration of sample N3 obtained by SEM reveals a graphene layer with numerous folds located on top of other intact layers that make up the stack. The observed folding was not induced by irradiation; however, it serves as evidence that connections between layers are weakened. Substantiating this hypothesis, the AFM-obtained image representing sample N3 displays leaf fragments overlaid on top of continuous flat layers.

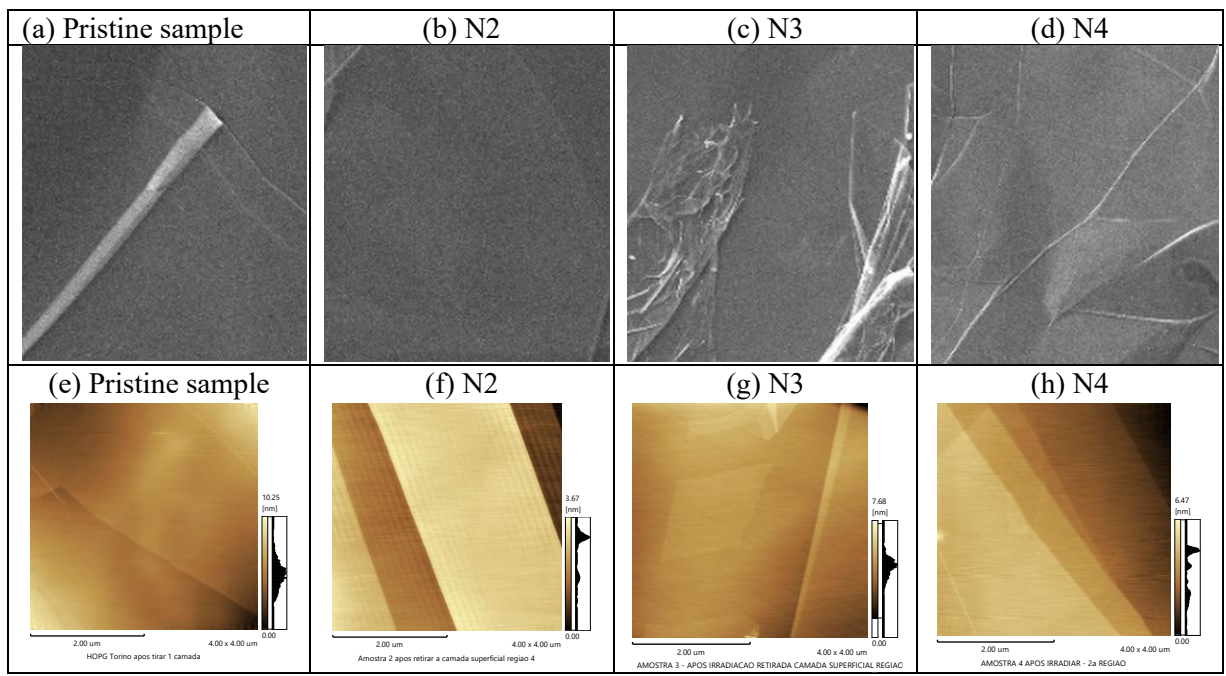

*Figure 8:* SEM micrograph [(a) to (d)] and Atomic Force Microscopy [(e) to (h)] images of the four HOPG foils irradiated with fast neutrons.

The RMS roughness and interlayer step distance values were 0.65(14) nm and 0.914(87) nm, respectively, for the pristine sample without irradiation. After exposure to the neutron beam for 30.5 h (sample N4), the RMS did not change. Nevertheless, the interlayer step distance value increased to 1.318(85) nm, which indicated a notable increase in the amount of loosely arranged planes, and this trend became more pronounced with extended radiation exposure time. This observation is evident in the images depicted in Figure 8.

## Conclusion

Simulations of the 14 MeV neutron reaction with $^{12}$C demonstrated that crystalline damage arises from various physical mechanisms, including atomic displacement damage and energy transfer to both the crystalline lattice and ionization of HOPG. The primary contribution to the degradation of the target's crystallinity is due to the interaction of the $^4$He particle, a product of the $^{12}$C$(n, \alpha)^9$Be reaction channel.

According to existing literature [13], as the crystalline perfection of HOPG shifts from a single crystal to a polycrystal and eventually to an amorphous state, thermal conductivity decreases by one order of magnitude at each phase. In this work, HOPG samples irradiated with different neutron fluence were studied using several techniques, and all the results indicated a loss in the crystalline perfection of HOPG. In the Raman spectra analysis, a consolidated D peak confirms the formation of point defects within the plane, manifesting as disordered regions in the crystalline lattice. The observed increase in intensity and broadening of the D peaks in the Raman results serve as additional evidence for the emergence of turbulence and disorder in the basal planes. Notably, despite these disturbances, the material retains its coherent layered graphite structure. Also, from the micro-Raman analysis, the gradual increase in the D/G intensity ratio indicated that the structural order of the materials was decreased [40].

Electrical resistivity, magnetoresistance, and thermal conductivity results showed a sharp drop in the values of these quantities. In particular, the thermal conductivity of HOPG was reduced between 52% (N3) and 59% (N2) and this can be attributed to the increased number of defects produced in the HOPG crystal lattice as a function of irradiation with fast neutrons, as indicated in reference [13]. The AFM and SEM indicated that the bond between graphene sheets weakened due to radiation exposure.

In conclusion, the effects of ionizing radiation result in structural defects in HOPG, generating vacancies within the graphene layers and interstitials between the layers, consequently reducing thermal conductivity. Even a small concentration of defects in the crystal lattice will likely cause a drastic reduction in this parameter. Considering that HOPG serves as the substrate for the films to be irradiated in the NUMEN Project, where the nuclear reactions of interest will occur, it is crucial to determine the decrease in thermal conductivity as a function of defect concentration. In the specific case of fabricating targets used in nuclear reactions, this information is vital for determining how frequently the target-substrate systems should be replaced during irradiation. Our findings will certainly help a more in-depth understanding of the effects of radiation, especially the tolerance of HOPG for nuclear applications.

## Acknowledgments

The authors acknowledge financial support from the funding agencies: INCT-FNA, CNPq proc. n. 464898/2014-5; proc. n. 303295/2022-8, proc. n. 408800/2021-6, proc. n. 44259/2019-6 and 304651/2021-4, FAPESP proc. n. 2023/16053-8, SisNANO/MCTI, Centro Universitário FEI and the NUMEN Project.